\documentclass[a4paper,UKenglish,cleveref, autoref]{lipics-v2019}
\usepackage{cite} 
\usepackage{url} 
\usepackage[utf8]{inputenc} 
\usepackage{booktabs} 
\usepackage{graphicx}
\usepackage[
    type={CC},
    modifier={by-nc-sa},
    version={4.0},
]{doclicense}
\nolinenumbers

\usepackage{xcolor}
\usepackage{tikz}
\newcommand{\circled}[1]{
        \setbox0=\hbox{#1}%
        \dimen0\wd0%
        \divide\dimen0 by 2%
        \begin{tikzpicture}[baseline=(a.base)]%
        \useasboundingbox (-\the\dimen0,0pt) rectangle (\the\dimen0,1pt);
        \node[circle,draw,outer sep=0pt,inner sep=0.1ex] (a) {#1};
        \end{tikzpicture}
}
\usepackage{listings}
\usepackage{float}
\newfloat{lstfloat}{htbp}{lop}
\floatname{lstfloat}{Listing}
 
\makeatletter
\newcommand\BeraMonottfamily{%
        \def\fvm@Scale{0.85}
        \fontfamily{fvm}\selectfont
}
\makeatother
\definecolor{eclipseBlue}{RGB}{42,0.0,255}
\definecolor{eclipseGreen}{RGB}{63,127,95}
\definecolor{eclipsePurple}{RGB}{127,0,85}
\lstdefinelanguage{sparql}{
        morecomment=[l][\color{olivegreen}]{\#},
        morestring=[b][\color{blue}]\",
        morekeywords={SELECT,CONSTRUCT,DESCRIBE,ASK,WHERE,FROM,NAMED,PREFIX,BASE,OPTIONAL,FILTER,GRAPH,LIMIT,OFFSET,SERVICE,UNION,EXISTS,NOT,BINDINGS,MINUS,a},
        sensitive=true
}

\usepackage[inline]{enumitem}
\usepackage{multirow,bigdelim}
\usepackage{booktabs}
\usepackage{graphics}
\usepackage{amsmath,amssymb,amsthm}
\usepackage{ifsym}
\usepackage{xfrac}
\usepackage{stmaryrd} 

\DeclareSymbolFont{largesymbolsA}{U}{jkpexa}{m}{n}
\SetSymbolFont{largesymbolsA}{bold}{U}{jkpexa}{bx}{n}
\DeclareMathSymbol{\varprod}{\mathop}{largesymbolsA}{16}
\usepackage{braket}

\usepackage{array}
\usepackage{arydshln}

\title{A framework supporting imprecise queries and data}
\author{Giacomo Bergami}{
Self-Conducted Project
}{bergamigiacomo@gmail.com}{https://orcid.org/0000-0002-1844-0851}{}
\acknowledgements{}
\Copyright{Giacomo Bergami}
\authorrunning{Giacomo Bergami}
\date{2019/07/02}
\hideLIPIcs

\begin{document}

\maketitle

\begin{abstract}
This technical report provides some lightweight introduction and some generic use case scenarios motivating the definition of a database supporting uncertainties in both queries and data.
This technical report is only providing the logical framework, which implementation is going to be provided in the final paper. 
\end{abstract}

\doclicenseThis

\section{Introduction}

\begin{figure}[!t]
        \centering
        \includegraphics[width=.7\textwidth]{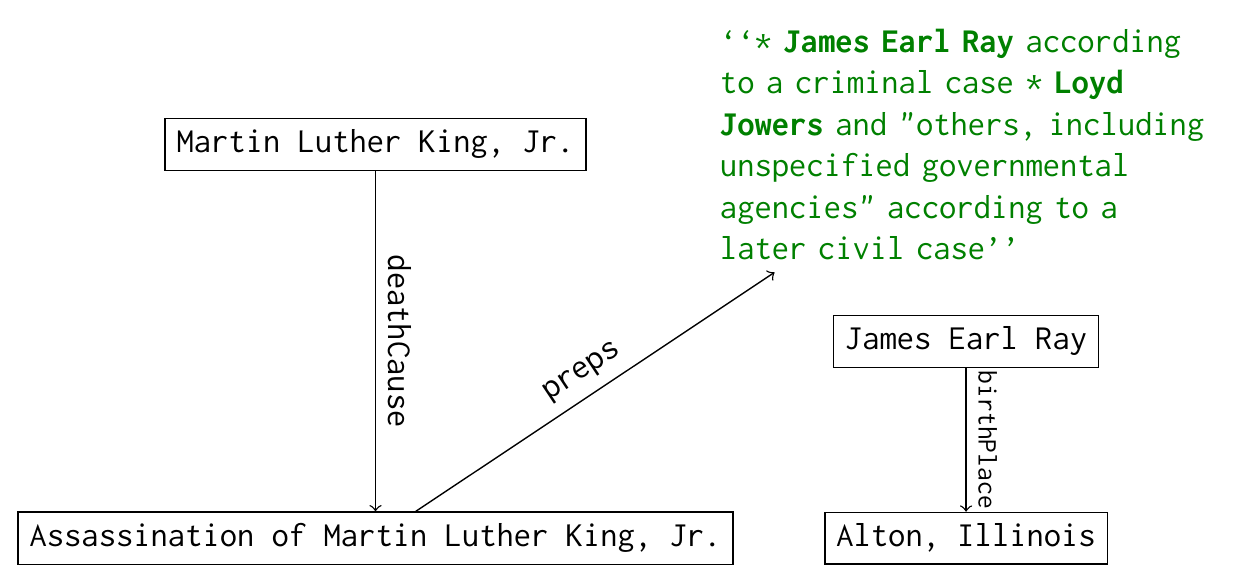}
        \caption{Answer relevant from the query in Listing \ref{lst:sparql} as retrieved from DBPedia. Please note that this answer contains disconnected information and approximately matches the query.}
        \label{ref:answer}
\end{figure}

\begin{lstfloat}[!b]
        \begin{lstlisting}[captionpos=b, caption=QALD-6 SPARQL query generated from the full-text ``\textit{In which town was the man convicted of killing Martin Luther King born?}'', label=lst:sparql,language=sparql]
PREFIX dbo: <http://dbpedia.org/ontology/> 
PREFIX dbp: <http://dbpedia.org/property/> 
PREFIX res: <http://dbpedia.org/resource/> 
SELECT DISTINCT ?uri WHERE {         
        res:Martin_Luther_King,_Jr. text:"the man convicted of killing King" ?x.         
        ?x dbo:birthPlace ?uri.
}
        \end{lstlisting}
\end{lstfloat} 
In industry ~\cite{amazon, google} as well as in academia ~\cite{nell, yago, reverb}, current knowledge base and linked data research focuses on automatic graph knowledge base construction and supporting efficient graph queries {(SPARQL or Cypher)} over graph databases {(RDF or Property Graph stores)}. In addition to these common usecases for graph databases, relational databases might be also be represented as graphs \cite{Chen2019}, and therefore graph queries might be also supported by relational databases \cite{SQLGraph}. Such knowledge bases  have been grown in popularity in several domain experts and decision makers such as governments and industrial stakeholders \cite{ArnaoutE18}: they are  interested in querying graph data  without necessarily knowing both the data schema and its representation \cite{YanMLC17}. As a result, the graph query generated from their full text information need  might have a significantly different schema from the one represented in the graph database and, still, the query answering system needs to provide neither a few nor empty answers \cite{YanMLC17}. In addition to this, full text imputation comes with also with user typos, thus resulting into entities that are hard (if not impossible) to match in the exact graph querying scenario \cite{ArnaoutE18}. In addition to that, automatic query generation tasks \cite{ZhengYZC18} might refuse to generate queries from full text if no exact linking was available \cite{ZhengYZC18}. An example of such scenario is provided by the QALD-6\footnote{\url{http://qald.sebastianwalter.org}} hybrid question answering dataset, and an example of such (SPARQL) query is provided in Listing \ref{lst:sparql}. As we can see, the query generation from full text might fail to link the text to the relationships represented within the knowledge base and, in the case the user inputs typos, it might also fail to link the entities to the ones in the knowledge base. As a result, no exact match interpretation of this generated query can return the expected result in Figure \ref{ref:answer}: at the time of the writing, no graph database system supports approximate graph matching for either imprecise data or imprecise queries. 
. In both approximate query answering and in answering a query having a different schema of the integrated knowledge base, the problem reduces in considering the query as the final hub schema \cite{HartungGR13}, over which aligning the data source to the query also corresponds to a graph matching phase.

Furthermore, Modern \textsc{Knowledge Base Management Systems} allow the ingestion of various data representations, such as relational, graph and full-text data \cite{ibmwatson,amsdottorato8348}. Current KBMS ingest  data sources that might be reliable as well as contain contradictory pieces of information. For the former specimen we might consider encyclopaedic data \cite{ibmwatson}, business data \cite{Saha16},  medical journals and clinical data \cite{WANG201834}, while for the latter we can include on-line social network tweets \cite{Lazer1094} or even medical diagnoses \cite{imihl}. Such inconsistency might be detected by linking \textbf{hierarchical datasets} such as ontologies \cite{mesh}, taxonomies \cite{icd11} and semantic networks \cite{Hoffart13,google,SpeerCH17} to the original data sources \cite{Lependu11}.

Despite there are several  databases and logical data models, there is currently no system that is able to  query a database where errors may be contained in both the query and the database instance itself. This paper is going to show how such challenge could be faced by generalizing well known approaches from the state of the art, as well as proposing novel contributions.

\section{Logical Framework}
This section provides which are the required constituent tools to implement a system matching all the aforementioned requirements. In this section, we give for granted that all the data sources have been integrated, so multiple instances of the same entity are grouped together within one same cluster: this step is fairly common in current literature \cite{Cebiric2018,BergamiBM19}, and therefore this preliminary step won't be detailed by the present technical report. Then, we need first to expand the knowledge base using some ad-hoc logical rules, that are going to draw inconsistencies. This grounding step needs to happen before query time (\S \ref{subsec:kbe}). After observing that querying data can be represented as a unification/grounding step of the query, we show how we can extend this approach to approximate graph matching, thus allowing to return approximated results when the right or expected information is not available within the knowledge base (\S\ref{subsec:agm}). As a side effect, this is going to provide a first hypothesis rank based on the query similarity. Last, we're going to rank all the hypotheses in a probabilistic fashion using the possible world enumeration over the finite possible worlds retrievable from the knowledge base (\S\ref{subsec:hwg}).

 \subsection{Grounding and Knowledge Base Expansion} \label{subsec:kbe}
\textbf{\textit{Preliminaries and Notation.}} In our scenario, the \textit{database} associated to the event knowledge base is the union of all the possible data sources $d_i$ represented as a set of tuples and describing different biased point of views potentially mutually inconsistent, that is $\mathbb{DB}_w=\bigcup_{d_i\in D}d_i$. Our database  is agnostic on the actual data provenance information. The database is also associated to a \textit{schema} $\mathcal{DB}=(\mathbf{DB},\Sigma_{DB})$, where $\mathbf{DB}$ is a finite set of relation symbols (atoms) and $\Sigma_{DB}$ is a finite set of constraints (or queries) over the atoms in $\mathbf{DB}$ \cite{Picado17}. Any hard constraint value in $\lambda\in \Sigma_{DB}$ has an associated weight $w(\lambda)\to +\infty$, and $[0,+\infty)$ if it is a soft constraint/rule. Each tuple $e$ in $\mathbb{DB}_w$ is associated to a confidence extraction weight $w(e)$ and is associated to a type $T\in \mathbf{DB}$. The database induces a finite domain of values $D=(\bigcup_{e\in \mathbb{DB}}\bigcup e)\backslash\{\texttt{NULL}\}$, thus obtaining a set of all the entities depicted in the relational database.

A knowledge base expansion over a relational database $\mathbb{DB}$ over a set of rules $\Sigma_{DB}$ is defined as the iterated application of the rules $\lambda \in \Sigma_{DB}$ over $\mathbb{DB}$ until no further knowledge is provided. Let us assume that $\mathbb{DB}$ also includes the hierarchical datasets describing the entities $D$ in their transitive and reflexive closure: such process can be described as a fixed point over $\mathbb{DB}$ via an unification algorithm. In particular, An \textit{unification function} $\theta$ is a function mapping a set of variables into a value in $D$. An \textit{unification algorithm} $U(\mathbb{DB},\lambda)$ for a rule $\lambda$ in the form of $\forall^h\exists^k\bigwedge_i B_i\Rightarrow H$ over the  database $\mathbb{DB}$ \cite{Martelli82} returns a set of unifications $\Theta$ such that each unification $\theta \in \Theta$ grounds an atom $B_i$ into an event of the database, that is $\theta B_i\in\mathbb{DB}$. A \textit{knowledge expansion step} (or ``grounding'') uses the unification algorithm to add new events that are generated from the rules' heads as follows:
\[ES_{\Sigma_{DB}}(\mathbb{DB}):=\mathbb{DB}\cup \Set{\theta H|\forall^h\exists^k\wedge_iB_i\Rightarrow H\in\Sigma_{DB},\theta\in U(\mathbb{DB},\lambda)}\]
Therefore, the \textit{knowledge expansion} algorithm is the least fix point of the $KE_{\Sigma_{DB}}$, that is: \[KE_{\Sigma_{DB}}(\mathbb{DB}):=\bigcup_{n\geq 0} ES_{\Sigma_{DB}}^n(\mathbb{DB})=\mathbb{GB}\]
where $\mathbb{GB}$ is the grounded knowledge base. This definition might be trivially extended to jointly perform 
contemporaneously the formulae grounding from $\Sigma$ and the factor graph generation over the finite model $\mathbb{DB}$, thus describing the knowledge expansion process.
\medskip
         
\textbf{\textit{Grounding for inconsistency detection.}}
At this stage, let us suppose that 
the query $\forall^h\exists^k\bigwedge_iB_i\Rightarrow H$ could be skolemized so that everything could be expressed only with the universal quantifier \cite{Harrison0022394}, and then the possibly bounded universal quantifiers could be  grounded with the actual $\mathbb{GB}$ values as a set of queries  $\mathcal{Q}=\{q_j:=\overline{B}_i^j\Rightarrow\overline{H}^j\}_{j\leq s, s\in \mathbb{N}}$. For the moment, we'll consider $q$ as just one of these $q_j\in\mathcal{Q}$ queries, and  we'll generalize the scenario by allowing with both  full-text,  multiple queries, and the approximate matching scenario in the next section. 

As we may observe, the actual knowledge base expansion part provides the actual reasoning part, where logical rules are actually applied and new knowledge is inferred. As a consequence of our observations in paraconsistent logic, the probabilistic reasoning will occur only after the actual logical inference as a way to rank all the possible words that could be inferred from one single query (\S\ref{subsec:hwg}). Consequently, the query answering system could directly work on the expanded knowledge base, and collect all the atoms matching with the query. As a result, all the possible worlds of interest that need to be ranked are only the possible subsets $\Braket{e_1,\dots,e_m}$ of $\mathbb{GB}$ matched by the query $q$ for returning the expected result $\gamma$; more formally, there exists an unification $\theta$ such that $\theta q=\wedge_{1\leq l\leq m}e_l\Rightarrow \gamma$. We will refer to such subsets of the grounded knowledge base as \textit{candidate hypothesis}.
$\gamma$ is then valid if and only if its expansion at any stage $n\geq 0$ generates no contradiction, $\bot$; therefore, it should be trivially true that $\forall \gamma. \gamma\neq \bot$. As a consequence, we need to also consider the queries in $\mathcal{Q}$ as rules required for the expansion step, so that we can ensure that the query answering system does not potentially generate further inconsistencies. Given that we cannot forecast all the possible $\mathcal{Q}$, we will perform the knowledge base expansion for $\mathcal{Q}$ after the one required for $\Sigma_{DB}$. In order to track down which expanded elements lead to a contradiction, we define a \textit{logical provenance function} $\kappa$.  In order to do so, $\kappa$ is initially defined over the non-grounded initial $\mathbb{DB}$ elements as:
        \[\kappa^0(u)=\begin{cases}
        \{\Braket{u}\} & u\in \mathbb{DB}\\
        \emptyset & \textup{oth.}\\
        \end{cases}\] 
        Still, we need to extend $\kappa^0$  for the stepwise expansion, thus requiring the following new knowledge expansion step function:
        \[{ES'}_{\Sigma_{DB}}(\Braket{\mathbb{DB},\kappa})=
        \Bigg\langle {ES}_{\Sigma_{DB}}(\mathbb{DB}),\quad
        \kappa\Big[\theta H\mapsto \kappa(\theta H)\cup \bigcup_{i=1}^n \kappa(\theta B_i)\Big]_{\substack{\bigwedge_{1\leq i\leq n}B_i\Rightarrow H=\lambda\in \Sigma_{DB} \theta \in U(\mathbb{DB},\lambda)}}\Bigg\rangle\\\]
        As a consequence, the resulting knowledge expansion algorithm could be defined via $ES'$ as follows:
        \[{KE'}_{\Sigma_{DB}}(\Braket{\mathbb{DB}, \kappa^0}):=\bigcup_{n\geq 0}{ES'}^n_{\Sigma_{DB}}(\Braket{\mathbb{DB}, \kappa^0})=\braket{\mathbb{GB}, \kappa^f}\]
        This fixed point produces a pair $\braket{\mathbb{GB}, \kappa^f}$, where $\kappa^f$ is the desired final function for all the possible expansion steps. The function $\kappa^f$ is also able to detect which elements are sufficient to draw a contradiction, thus obtaining the overall \textit{minimal inconsistent subsets} as in \cite{Hunter08}  by calling $\kappa^f(\bot)$. In particular, we can check if any possible subset of $\mathbb{GB}$ is inconsistent using the following predicate using $\kappa^f$:
\[\texttt{Cons}(\vec{e}):=\neg\exists e'\in \kappa^f(\bot). e'\subseteq \vec{e} \]

Therefore, the set of the valid events generating hypotheses for candidate $\gamma\neq\bot$ from an answered query $q$ is defined as all the possible subsets of the grounded knowledge base over $ {KE'}_{\Sigma_{DB}\cup\mathcal{Q}\cup{A}(\mathcal{Q},\mathcal{DB})}(\Braket{\mathbb{GB},\kappa^f})$ that do not contain minimal inconsistent subsets. Given that approximate graph matching requires aligning multiple relationships into possible one single relationship \cite{Hu0YWZ18}, hen we need to infer such expansion or reduction rules from an aligning task ${A}$ as extended in \cite{amsdottorato8348} generating the correspondences for matching the set of queries in $\mathcal{Q}$ (source schema/ontology) to the knowledge base schema $\mathcal{DB}$ (target schema/ontology). With an abuse of notation introduced not to further complicate the report's notation, we denote the output of this further expansion step also as $\Braket{\mathbb{GB},\kappa^f}$. Therefore, the set of the consistent unifications for the query $q$ is given as follows:
\[\mathcal{U}_{\mathbb{GB}}(q)=\Set{\wedge_{1\leq l\leq m}e_l\Rightarrow \gamma | \theta\in U(\mathbb{GB},q).\theta q=\wedge_{1\leq l\leq m}e_l\Rightarrow \gamma \wedge \texttt{Cons}(e_1,\dots,e_m)}\]
$\mathcal{U}_{\mathbb{GB}}(q)$ does not return set of events generating contradictions by construction. The requirement and assumption that  $\gamma\neq\bot$ does not break the generality of the querying approach: in fact,  the user could still query which are all the (minimal) inconsistent hypotheses by invoking $\kappa^f(\bot)$, thus allowing to get all the possible MIS in a tractable time, that is linearly to the number of the expansion steps required to ground the knowledge base which, by previous assumptions, is polynomial over the size of the data.
At the moment, we won't consider any paraconsistent reasoning to tolerate inconsistent or incomplete queries with respect to the query's completeness; we'll focus more on these aspects in the next section, where approximated queries will be introduced.
On the other hand, this definition also guarantees that each generated candidate has associated some non-empty evidence supporting it, that is it is always associated to a non-empty set of hypotheses. We can show that in the following lemma:

We can then define a function $\mathcal{A}$ generating all the candidates $\gamma_k$ and a function $\mathcal{H}_{\mathbb{GB}}(q,\gamma_k)$ generating all the consistent candidates for $\gamma_k$.
\[\mathcal{A}^{\mathcal{U}}(\mathbb{GB},q):=\Set{\gamma | H\Rightarrow \gamma \in \mathcal{U}_\mathbb{GB}(q)}\qquad \mathcal{H}^{\mathcal{U}}_{\mathbb{GB}}(q,\gamma):=\tilde{\Gamma}(\Set{H|H\Rightarrow\gamma\in \mathcal{U}_\mathbb{GB}(q)})\]

where $\tilde{\Gamma}$ is an hypothesis summarization function using a similarity function $\sim$ to merge similar results. At the moment we shall consider $\sim$ as the equivalence similarity function $\equiv$ for which $\forall S. \overset{\equiv}{\Gamma}(S)=S$.  
When the algorithm ${\mathcal{U}}$ used for generating the expansions from the consistent unification, then we shall write $\mathcal{A}$ (or $\mathcal{H}$) instead of $\mathcal{A}^\mathcal{U}$ (or $\mathcal{H}^\mathcal{U}$). A dot $\cdot$ replaces $\mathcal{U}$ when we want to express that any unification algorithm $\mathcal{U}$ could be used instead (e.g., $\mathcal{A}^\cdot$). After providing those definitions, we can now guarantee that if $\mathcal{H}$ returns no hypotheses associated to the target $\gamma$ for query $q$, then it means that either no exact and consistent subset of  matched elements from the grounded knowledge base is possible.

\begin{lemma}\label{firstlemma}
\[\mathcal{H}^{\mathcal{U}}_{\mathbb{GB}}(q,\gamma)=\emptyset\Leftrightarrow \forall \theta\in U(\mathbb{GB},q). \theta q=\theta B\Rightarrow \gamma \wedge \theta B\in \mathbb{GB}\wedge \exists e\in \kappa^f(\bot). e\subseteq\mathbb{GB}\]
\end{lemma}
\begin{proof}
We can rewrite the following assertion by equivalence as follows:
\[\begin{split}
\mathcal{H}^{\mathcal{U}}_{\mathbb{GB}}(q,\gamma)=\emptyset\Leftrightarrow\neg\exists H. H\Rightarrow\gamma \in\mathcal{U}_{\mathbb{GB}}(q)\Leftrightarrow\forall H.H\Rightarrow\gamma\notin \mathcal{U}_{\mathbb{GB}}(q)
\end{split}\]
The lemma is then proved by expansion of the definition of $\mathcal{U}$: in fact, $\mathcal{U}$ does not contain a rule $H\Rightarrow\gamma$ if and only if there exists any possible unification $\theta$ over $\mathbb{GB}$ of $q$ that produces that grounded rule, or if all the possible unifications $\theta$ produce a head which is inconsistent, e.g., $\neg\texttt{Cons}(H)$. 
\end{proof}
This proof should be further enforced when we will replace the exact unification algorithm with an approximated one in \S\ref{subsec:agm}.

Finally, the set of all the answer associated to all the possible worlds as in Table \ref{fig:6} could be represented by the following \textit{justification} function:
\[\forall \gamma \in\mathcal{A}^\mathcal{U}(\mathbb{GB},q).\quad \mathcal{J}^\mathcal{U}_{\mathbb{GB},q}(\gamma):=\mathcal{H}^\mathcal{U}_\mathbb{GB}(q,\gamma)\]
Even in this case, we shall write $\mathcal{J}$ instead of $\mathcal{J}^\mathcal{U}$ when $\mathcal{U}$ is clear from the context.
\subsection{Approximate Graph matching}\label{subsec:agm}

As previously discussed, this phase is the one substituting the traditional exact query matching in relational databases and providing the approximation for the query answering scenario. We want also to remark why the graph query answering and approximate matching problem is the most general possible solution for a query answering scenario, thus also including the full-text query answering. 
\medskip

\textbf{\textit{Problem statement and generalization.}} Let us suppose that $q$ is a  full-text query: an graph database and  domain expert might ``compile'' the information need expressed by $q$ as a  domain-specific gold query $q^G$,  perfectly matching its informative needs. On the other hand, the real-world domain expert is does not generally know how to query a graph \cite{FaderZE14} (or, as in this case, a relational \cite{LiJ14}) database, and therefore uses an automated compiler $\mathcal{C}\colon D\to \wp(\mathcal{L})$ to translate each full text query $d\in D$ into multiple possible queries \cite{LiJ14}. Each generated query $q_i\in\mathcal{L}$ also comes with an interpretation error $\varepsilon(q_i)\in[0,1]$ \cite{LiJ14}. 

Let us suppose to have a graph similarity function $\sim$ as the one initially discussed for graph matching in the related work section (i.e., where both structural and semantical features might be combined). We are now ideally   able to measure $\varepsilon(q_i)$ as $1-(q^G\sim q_i)$, that is how similar is the expected graph query to the one that has been generated by the graph query compiler $\mathcal{C}$. On the other hand, $q^G$ is not generally available, and therefore the only way to assess this error is to evaluate $1-(g(q)\sim q_i)$ instead, where $g(q)$ is the graph representation of the full text query \cite{amsdottorato8348}. Consequently, the $\sim$ (or its opposite, $\varepsilon$) function should be already used by $\mathcal{C}$ while generating the queries from the full-text, so that the quality of the query generation output could be assessed.
\medskip

The generality of this compiler framework for the full-text interpretation scenario could be assessed by looking at all the different possible implementations of $\mathcal{C}$ depending of the querying scenario of interest:
\begin{enumerate}
\item when $q$ is a query that needs to be aligned towards different possible knowledge base schemas, then  each different query $q_i$ represents $q$ for a specific knowledge base schema;
\item when $q$ is a skolemized FOL query belonging to the AE fragment, then $q_i$ represents each possible combination of variables grounded for the bounded universal quantifiers. 
\item when $q$ is a combinatio of all the three aforementioned query scenarios, the associated compiler is a specific combination of the aforementioned query compilers.
\end{enumerate}

In this generalization step, we can then generalize $\mathcal{A}$, $\mathcal{H}$ and $\mathcal{J}$ by simply adopting a new generalization algorithm $\mathcal{U}^\mathcal{C}$ (\textit{compiler generalization}), which is defined as follows:
\[\mathcal{U}^\mathcal{C}(\mathbb{GB},q)=\bigcup_{q_i\in\mathcal{C}(q)}\mathcal{U}(\mathbb{GB},q)\]

\medskip

\textbf{\textit{Result generation.}} By extending the definition of graph query languages, approximate graph query answering systems such as \cite{DeVirgilio2015} also provide a set of subgraphs (or morphisms) as the result of the graph matching phase. Therefore, given a graph database/knowledge graph $\mathbb{DB}$, $q_i(\mathbb{DB})$ returns a set of graphs $\{\tilde{g}_j\}_{j\leq n, n\in \mathbb{N}}$, where the same $\sim$ function -- also used by the compiler --  is used to define the query result ranking function. The requirement that $\sim$ is kept the same throughout the whole approximate querying process is crucial, as it guarantees that we're always judging and ranking the data in the same way, thus allowing better optimization tasks. 

Within the query generation and decomposition task for approximate graph matching outlined in \cite{DeVirgilio2015}, in addition to structural clusters we might want to adopt semantic clusters as well: that is always possible via the unique metric $\sim$ that is used throughout the process. After the graph combination phase that could be carried out in a multi-way join fashion as outlined in \cite{Chen2019}, we might consequently obtain graph result clusters via the combination of the intermediate subgraph clusters, thus already subsets of different solutions. These representations of graph collections in such clusters can be then further summarized as one single graph using graph summarization algorithms \cite{Cebiric2018,BergamiBM19}: in particular, we require a clustering and summarization function   $\tilde{\Gamma}$ that, taken as an input a collection of graphs, it will indirectly use the similarity function $\sim$ for the clustering and then summarizes the clusters using an UDF function. 
Given $\nu_{\textrm{UDF}}$ the graph summarization function \cite{BergamiPM18}, we can then define $\tilde{\Gamma}$ as follows:
\[\tilde{\Gamma}(G)=\Set{\nu_{\textrm{UDF}}(\Set{g'\in G|g\sim g'>\theta_r}) | g\in G}\] where $\theta_r$ is a similarity threshold value identifying if the hypotheses are similar or not (see \S\ref{subsec:agm}). Given that the unification algorithms guarantee the consistency of the result by returning only consistent results, nothing is guaranteed that the cluster of similar graphs are indeed also (at least partially) consistent between each other. In order to provide such guarantee, we should impose that all the similarity functions $\sim$ provide such guarantee:
\[g\sim g'=1\Rightarrow \mathtt{Cons}(g\cup g')\]
Please note that it is not possible to evaluate the opposite direction of the logical implication, because we might also have two graphs $g$ and $g'$ that are consistent but not necessarily similar or identical.

Given that the $\nu$ operation presented in \cite{BergamiPM18} preserves the original graph $g\in\ G$ information, we can always track each summarized graph to all the graphs providing such summarization and define the following \textit{graph provenance} function:
\[\forall s\in\tilde{\mathcal{J}}_{\mathbb{GB},q}^\mathcal{C}(\gamma).\quad \tilde{\mathcal{J}}\mathbf{map}\;^{\mathcal{C}}_{\mathbb{GB},q}(s):={\max\arg}_{G'\subseteq \mathcal{J}_{\mathbb{GB},q}^\mathcal{C}(\gamma)}\;\tilde{\Gamma}(G')=\{s\}\]

As a last requirement, $\sim$ shall also consider inconsistency detection via $\Sigma_{DB}$, such that all the possible subgraphs of $\bigcup \tilde{\mathcal{J}}\mathbf{map}\;^{\mathcal{C}}_{\mathbb{GB},q}(s) $ are not in $\kappa^f(\bot)$, thus requiring that the summarization of similar graph does not lead to an inconsistent graph (i.e., hypothesis).

\subsection{Hypothesis World Generation}\label{subsec:hwg}

As stated in the introduction, our decision support system does not want to  provide  \textbf{the} correct answer to a query, but only ranking all the provided results by data reliability and query matching. 
Therefore, we don't want to evaluate the probability associated to each candidate answer $\gamma_k$ in comparison with all the other remaining ones, but the probability of $\gamma_k$ assuming that it might be a correct answer to our problem. More formally, we require instead that each $\gamma_k$ has probability $\mathbb{P}(\gamma_k)=1$ if and only if all both  the candidate hypotheses contain no data uncertainties and all the candidate hypotheses perfectly match with the query. 
 For each $\gamma_k$ we are  only interested on the possible worlds that are generated by the query evaluation of $q$ and the associated grounding process. 
All the other remaining worlds  must be considered as irrelevant w.r.t. the candidate, and therefore with zero probability. 

In order to assess such probability values, we can choose to model our grounded knowledge base as a Markov Logic Networks model using \textit{factor graphs}. Each $F_i$, either an atom $A$ or a  grounded formula $\theta H \leftarrow\wedge_i\;\theta B_i$, has an associated exponential factor $\phi_i\in \Phi$, which domain $X_i$ is  the set of the grounded atoms appearing in $F_i$. A factor graph over a set of factors $\Phi$ for a grounded knowledge base $\mathbb{GB}$ induces the following joint probability distribution of stochastically independent worlds:
\[\mathbb{P}(\vec{X}=\vec{x})=\sfrac{1}{Z} \prod_{\phi_i\in \Phi}\phi({X}_i)^{n_i({X}_i)}\]
where $Z$ is the normalization  over the allowed possible worlds $\mathcal{W}$   in the universe $\mathcal{U}$ such that $\sum_{\mathcal{W}\in \mathcal{U}}\mathbb{P}(\vec{X}=\mathcal{W})=1$, and $n_i({X}_i)$ is the number f true groundings $X_i$ of each  $F_i$ associated to  $\phi_i$. Each possible world $\mathcal{W}$ is a truth assignment over the atoms in $\vec{X}$, while $\mathcal{U}$ is defined as a finite partition of a sample space represented by the union of all the possible worlds.

The following lemma shows that such graph model includes as  irrelevant the worlds containing contradictions:

\begin{lemma}
        Given a factor graph described by the set of factors $\Phi$ describing both the database atoms and the grounded rules, all the worlds containing a contradiction have zero probability.
\end{lemma}
\begin{proof}
        Without any loss of generality, let us suppose that our factor graph contains only two grounded atoms $A$ and $B$ and a grounded hard rule with infinite weight stating that such atoms are contradictory, and therefore $F=A\wedge B\to \bot$; therefore, we have three factors $\phi_i(\vec{x}_i)$, both for the atoms and for the grounded formula. Our model is $\mathbb{P}(A,B)=\sfrac{\phi_A(A)\phi_B(B)\phi_F(A,B)}{Z}$, where the normalization factor $Z$ is defined as $e^{\theta_A+\theta_B}+e^{\theta_F}(1+\theta_A+\theta_B)$. For the only contradictory world where both $A$ and $B$ hold we have that $\mathbb{P}(1,1)=\left(1+e^{\theta_F}\cdot \frac{1+e^{\theta_A}+e^{\theta_B}}{e^{\theta_A+\theta_B}}\right)^{-1}$, such that for $\theta_F\to+\infty$ has a zero probability.
\end{proof}

In order to meet the requirement set up in this subsection's first paragraph, we define an universe $\mathcal{U}_{\gamma_k}$ for each possible candidate ${\gamma_k}$, where each worlds is to be identified within the graph $\bigcup\tilde{\mathcal{J}}^{\mathcal{C}}_{\mathbb{GB},q}({\gamma_k})$: given that the resulting graph could be a summarized version of $\bigcup{\mathcal{J}}^{\mathcal{C}}_{\mathbb{GB},q}({\gamma_k})$, each \textit{candidate hypothesis} should be evaluated starting from $\bigcup\tilde{\mathcal{J}}\mathbf{map}\;^{\mathcal{C}}_{\mathbb{GB},q}(s)$ for each $s\in\tilde{\mathcal{J}}^{\mathcal{C}}_{\mathbb{GB},q}({\gamma_k}) $. Let us remember that, by construction, each candidate hypothesis never contains inconsistencies. Given that we are generally not interested in all the possible sub-worlds containing $\bigcup\tilde{\mathcal{J}}\mathbf{map}\;^{\mathcal{C}}_{\mathbb{GB},q}(s)$ and that it is more interesting to see which is the biggest relevant and consistent world containing it, we shall define the maximum relevant world where $s$ holds as:
\[\llbracket s \rrbracket_q^{\sim}:=\bigcup_{\vec{e}\in\tilde{\mathcal{J}}\mathbf{map}\;^{\mathcal{C}}_{\mathbb{GB},q}(s)}\Set{e\in \mathbb{GB}|\exists c\in \kappa^f(e).\; c\subseteq\vec{e}}\]

Therefore, we have that the universe where ${\gamma_k}$ holds with relevance to the query is $\mathcal{U}_{\gamma_k}:=\Set{\llbracket s \rrbracket_q^\sim|  s\in\tilde{\mathcal{J}}_{\mathbb{GB},q}^\mathcal{C}({\gamma_k})}$. Finally, we can express the probability of the candidate $\gamma$ subordinate to the data extraction errors by using the law of total probability as in  \cite{doi101198}, that is $\mathbb{P}({\gamma_k};w):=\sum_{\mathcal{W}\in\mathcal{U}_{\gamma_k}}\mathbb{P}(\mathcal{W}\in \mathcal{U}_{\gamma_k}|w(\mathcal{W}))\mathbb{P}(\mathcal{W})$: $\mathbb{P}(\mathcal{W})$ is the usual joint probability distribution induced by the factor graph considering the data extraction errors, and  the likelihood $\mathbb{P}(\mathcal{W}\in \mathcal{U}_{\gamma_k}|w(\mathcal{W}))$ can be represented by a scoring function combining the query $q_i$ similarities that generated the results as well as the similarity of each matched subgraph with respect to the query.
For efficiency reasons, such score $w$ shall be evaluated during the $\tilde{\Gamma}$ operation.
        \bibliographystyle{abbrv}
        \bibliography{sigproc}

\begin{thebibliography}{10}

\bibitem{ArnaoutE18}
H.~Arnaout and S.~Elbassuoni.
\newblock Effective searching of {RDF} knowledge graphs.
\newblock {\em J. Web Sem.}, 48:66--84, 2018.

\bibitem{amsdottorato8348}
G.~Bergami.
\newblock {\em A new Nested Graph Model for Data Integration}.
\newblock PhD thesis, Alma Mater Studiorum -- University of Bologna, 2018.

\bibitem{BergamiBM19}
G.~Bergami, F.~Bertini, and D.~Montesi.
\newblock On approximate nesting of multiple social network graphs: a
  preliminary study.
\newblock In {\em Proceedings of the 23rd International Database Applications
  {\&} Engineering Symposium, {IDEAS} 2019, Athens, Greece, June 10-12, 2019.},
  pages 40:1--40:5, 2019.

\bibitem{BergamiPM18}
G.~Bergami, A.~Petermann, and D.~Montesi.
\newblock Thosp: an algorithm for nesting property graphs.
\newblock In A.~Arora, A.~Bhattacharya, G.~H.~L. Fletcher, J.~Larriba{-}Pey,
  S.~Roy, and R.~West, editors, {\em Proceedings of the 1st {ACM} {SIGMOD}
  Joint International Workshop on Graph Data Management Experiences {\&}
  Systems {(GRADES)} and Network Data Analytics (NDA), Houston, TX, USA, June
  10, 2018}, pages 8:1--8:10. {ACM}, 2018.

\bibitem{Cebiric2018}
{\v{S}}.~{\v{C}}ebiri{\'{c}}, F.~Goasdou{\'e}, H.~Kondylakis, D.~Kotzinos,
  I.~Manolescu, G.~Troullinou, and M.~Zneika.
\newblock Summarizing semantic graphs: a survey.
\newblock {\em The VLDB Journal}, Dec 2018.

\bibitem{Chen2019}
Y.~Chen, B.~Guo, and X.~Huang.
\newblock $\delta$-transitive closures and triangle consistency checking: a new
  way to evaluate graph pattern queries in large graph databases.
\newblock {\em The Journal of Supercomputing}, Feb 2019.

\bibitem{imihl}
D.~Costa and M.~A. Martins.
\newblock {\em Measuring Inconsistency in Information}, chapter Inconsistency
  Measures in Hybrid Logics, pages 169--194.
\newblock College Publications, 2018.

\bibitem{DeVirgilio2015}
R.~De~Virgilio, A.~Maccioni, and R.~Torlone.
\newblock Approximate querying of rdf graphs via path alignment.
\newblock {\em Distributed and Parallel Databases}, 33(4):555--581, Dec 2015.

\bibitem{reverb}
A.~Fader, S.~Soderland, and O.~Etzioni.
\newblock Identifying relations for open information extraction.
\newblock In {\em Proceedings of the Conference of Empirical Methods in Natural
  Language Processing ({EMNLP} '11)}, Edinburgh, Scotland, UK, July 27-31 2011.

\bibitem{FaderZE14}
A.~Fader, L.~Zettlemoyer, and O.~Etzioni.
\newblock Open question answering over curated and extracted knowledge bases.
\newblock In {\em The 20th {ACM} {SIGKDD} International Conference on Knowledge
  Discovery and Data Mining, {KDD} '14, New York, NY, {USA} - August 24 - 27,
  2014}, pages 1156--1165, 2014.

\bibitem{google}
Google.
\newblock Freebase data dumps.
\newblock \url{https://developers.google.com/freebase/data}.

\bibitem{Harrison0022394}
J.~Harrison.
\newblock {\em Handbook of Practical Logic and Automated Reasoning}.
\newblock Cambridge University Press, 2009.

\bibitem{HartungGR13}
M.~Hartung, A.~Gro{\ss}, and E.~Rahm.
\newblock Composition methods for link discovery.
\newblock In V.~Markl, G.~Saake, K.~Sattler, G.~Hackenbroich, B.~Mitschang,
  T.~H{\"{a}}rder, and V.~K{\"{o}}ppen, editors, {\em Datenbanksysteme
  f{\"{u}}r Business, Technologie und Web (BTW), 15. Fachtagung des
  GI-Fachbereichs "Datenbanken und Informationssysteme" (DBIS), 11.-15.3.2013
  in Magdeburg, Germany. Proceedings}, volume 214 of {\em {LNI}}, pages
  261--277. {GI}, 2013.

\bibitem{Hoffart13}
J.~Hoffart, F.~M. Suchanek, K.~Berberich, and G.~Weikum.
\newblock Yago2: A spatially and temporally enhanced knowledge base from
  wikipedia.
\newblock {\em Artif. Intell.}, 194:28--61, Jan. 2013.

\bibitem{Hu0YWZ18}
S.~Hu, L.~Zou, J.~X. Yu, H.~Wang, and D.~Zhao.
\newblock Answering natural language questions by subgraph matching over
  knowledge graphs.
\newblock {\em {IEEE} Trans. Knowl. Data Eng.}, 30(5):824--837, 2018.

\bibitem{Hunter08}
A.~Hunter and S.~Konieczny.
\newblock Measuring inconsistency through minimal inconsistent sets.
\newblock In {\em Proceedings of the Eleventh International Conference on
  Principles of Knowledge Representation and Reasoning}, KR'08, pages 358--366.
  AAAI Press, 2008.

\bibitem{Lazer1094}
D.~M.~J. Lazer, M.~A. Baum, Y.~Benkler, A.~J. Berinsky, K.~M. Greenhill,
  F.~Menczer, M.~J. Metzger, B.~Nyhan, G.~Pennycook, D.~Rothschild,
  M.~Schudson, S.~A. Sloman, C.~R. Sunstein, E.~A. Thorson, D.~J. Watts, and
  J.~L. Zittrain.
\newblock The science of fake news.
\newblock {\em Science}, 359(6380):1094--1096, 2018.

\bibitem{Lependu11}
P.~Lependu and D.~Dou.
\newblock Using ontology databases for scalable query answering, inconsistency
  detection, and data integration.
\newblock {\em J. Intell. Inf. Syst.}, 37(2):217--244, Oct. 2011.

\bibitem{LiJ14}
F.~Li and H.~V. Jagadish.
\newblock Constructing an interactive natural language interface for relational
  databases.
\newblock {\em {PVLDB}}, 8(1):73--84, 2014.

\bibitem{yago}
F.~Mahdisoltani, J.~Biega, and F.~M. Suchanek.
\newblock {YAGO3:} {A} knowledge base from multilingual wikipedias.
\newblock In {\em {CIDR} 2015, Seventh Biennial Conference on Innovative Data
  Systems Research, Asilomar, CA, USA, January 4-7, 2015, Online Proceedings}.

\bibitem{Martelli82}
A.~Martelli and U.~Montanari.
\newblock An efficient unification algorithm.
\newblock {\em ACM Trans. Program. Lang. Syst.}, 4(2):258--282, Apr. 1982.

\bibitem{nell}
T.~Mitchell, W.~Cohen, E.~Hruschka, P.~Talukdar, J.~Betteridge, A.~Carlson,
  B.~Dalvi, M.~Gardner, B.~Kisiel, J.~Krishnamurthy, N.~Lao, K.~Mazaitis,
  T.~Mohamed, N.~Nakashole, E.~Platanios, A.~Ritter, M.~Samadi, B.~Settles,
  R.~Wang, D.~Wijaya, A.~Gupta, X.~Chen, A.~Saparov, M.~Greaves, and
  J.~Welling.
\newblock Never-ending learning.
\newblock In {\em Proceedings of the Twenty-Ninth AAAI Conference on Artificial
  Intelligence (AAAI-15)}, 2015.

\bibitem{mesh}
U.~N.~L. of~Medicine.
\newblock Mesh (https://www.nlm.nih.gov/mesh/).

\bibitem{ibmwatson}
I.~J. of~Research and Development.
\newblock {\em This is Watson}, volume 56(3/4).
\newblock IBM Co., May/July 2012.

\bibitem{Picado17}
J.~Picado, A.~Termehchy, A.~Fern, and P.~Ataei.
\newblock Schema independent relational learning.
\newblock In {\em Proceedings of the 2017 ACM International Conference on
  Management of Data}, SIGMOD '17, pages 929--944, 2017.

\bibitem{Saha16}
D.~Saha, A.~Floratou, K.~Sankaranarayanan, U.~F. Minhas, A.~R. Mittal, and
  F.~\"{O}zcan.
\newblock Athena: An ontology-driven system for natural language querying over
  relational data stores.
\newblock {\em Proc. VLDB Endow.}, 9(12):1209--1220, Aug. 2016.

\bibitem{doi101198}
N.~D. Singpurwalla and J.~M. Booker.
\newblock Membership functions and probability measures of fuzzy sets.
\newblock {\em Journal of the American Statistical Association},
  99(467):867--877, 2004.

\bibitem{SpeerCH17}
R.~Speer, J.~Chin, and C.~Havasi.
\newblock Conceptnet 5.5: An open multilingual graph of general knowledge.
\newblock In {\em Proceedings of the Thirty-First {AAAI} Conference on
  Artificial Intelligence, February 4-9, 2017, San Francisco, California,
  {USA.}}, pages 4444--4451, 2017.

\bibitem{SQLGraph}
W.~Sun, K.~Srinivas, A.~Fokoue, A.~Kementsietsidis, G.~Hu, and G.~Xie.
\newblock Sqlgraph: An efficient relational-based property graph store.
\newblock In {\em SIGMOD}, 2015.

\bibitem{amazon}
R.~Trivedi, B.~Sisman, X.~L. Dong, C.~Faloutsos, J.~Ma, and H.~Zha.
\newblock Linknbed: Multi-graph representation learning with entity linkage.
\newblock In I.~Gurevych and Y.~Miyao, editors, {\em Proceedings of the 56th
  Annual Meeting of the Association for Computational Linguistics, {ACL} 2018,
  Melbourne, Australia, July 15-20, 2018, Volume 1: Long Papers}, pages
  252--262. Association for Computational Linguistics, 2018.

\bibitem{WANG201834}
Y.~Wang, L.~Wang, M.~Rastegar-Mojarad, S.~Moon, F.~Shen, N.~Afzal, S.~Liu,
  Y.~Zeng, S.~Mehrabi, S.~Sohn, and H.~Liu.
\newblock Clinical information extraction applications: A literature review.
\newblock {\em Journal of Biomedical Informatics}, 77:34 -- 49, 2018.

\bibitem{icd11}
W.H.O.
\newblock Icd-11 (https://icd.who.int/).

\bibitem{YanMLC17}
L.~Yan, R.~Ma, D.~Li, and J.~Cheng.
\newblock {RDF} approximate queries based on semantic similarity.
\newblock {\em Computing}, 99(5):481--491, 2017.

\bibitem{ZhengYZC18}
W.~Zheng, J.~X. Yu, L.~Zou, and H.~Cheng.
\newblock Question answering over knowledge graphs: Question understanding via
  template decomposition.
\newblock {\em {PVLDB}}, 11(11):1373--1386, 2018.

\end{thebibliography}

\end{document}